\documentclass[nofootinbib,amsmath,amssymb,aps,floatfix,notitlepage,
superscriptaddress,
longbibliography]{revtex4-2}
\pdfoutput=1

\usepackage{graphics}
\usepackage{graphicx}
\usepackage{color}

\usepackage{enumitem}

\usepackage{amssymb}

\usepackage{hyperref}\hypersetup{colorlinks=true,linktoc=all,}

\def\hhref#1{\href{http://arxiv.org/abs/#1}{arXiv:#1}}

\begin{document}

\title{Borel Summation and Analytic Continuation of the  Heat Kernel on Hyperbolic Space}


\author{Gerald V. Dunne}

\affiliation{Department of Physics, University of Connecticut, Storrs CT 06269-3046}

\begin{abstract}
The heat kernel expansion on even-dimensional hyperbolic spaces is asymptotic at both short and long times, with interestingly different Borel properties for these short and long time expansions. Resummations in terms of incomplete gamma functions provide accurate extrapolations and analytic continuations, relating the heat kernel to the Schr\"odinger kernel, and the heat kernel on hyperbolic space to the heat kernel on spheres. For the diagonal heat kernel there is also a duality between short and long times which mixes the scalar and spinor heat kernels.

\end{abstract}

\maketitle

\centerline{\it Dedicated to Peter Suranyi,  whose breadth of knowledge and interests is an inspiration.}

\section{Introduction}
\label{sec:intro}

Heat kernels on hyperbolic space provide a fascinating arena for the study of the relation between spectral and geometric features of manifolds \cite{mckean,cartier,voros,davies,ottewill,Camporesi:1990wm,sarnak,grigoryan}. The long time expansion is naturally related to spectral analysis on the manifold, while the short time expansion is naturally related to a geometric expansion in terms of geodesics. Physically this corresponds to a Hamiltonian or a Lagrangian formulation of the heat propagation problem, respectively. Here we investigate heat kernels on even dimensional hyperbolic space, for which both the short time and long time expansions are asymptotic. 
Here we study these asymptotic expansions using Borel techniques, pointing out the similarities and differences. 

Hyperbolic spaces of constant curvature are harmonic spaces, and the heat kernel  $K(t, \rho)$ is a function of just  the elapsed time $t$ and the geodesic distance $\rho$ between the initial and final points. A simple recurrence relation \cite{davies,grigoryan} connects the expressions in dimension $d$ and dimension $d+2$: 
\begin{eqnarray}
K_{d+2}(t, \rho) = -\frac{e^{-d\, t}}{2\pi \, \sinh(\rho)} \frac{\partial}{\partial\rho} K_{d}(t, \rho)
\label{eq:recurrence}
\end{eqnarray}
This effectively reduces the asymptotic problem to the separate analysis of the $d=1$ and $d=2$ cases. The physical difference between odd and even dimensional spaces is related to Huygens' principle \cite{ooguri}. For odd dimensional spaces the heat kernel expansions are convergent and have a simple structure, which can be derived from the 1-dimensional case. On the other hand, for even dimensional spaces the heat kernel expansions are divergent and have a rich structure, which can be derived from the 2-dimensional case \cite{mckean}. Thus, we focus here on the heat kernel on $\mathbb H^2$, the two dimensional hyperbolic space of constant negative curvature, for which the short time and long time expansions are asymptotic. 
Since these expansions are asymptotic, interesting features arise in the analytic continuation of the heat kernel to the Schr\"odinger kernel ($t\to i\, t$), and of the heat kernel on hyperbolic space to the heat kernel on spheres (effectively $t\to -t$). An interesting feature in the short time expansion is the role of {\it complex} geodesics in the analytic continuation from hyperbolic spaces to spheres.
We also identify a duality ($t\to \frac{4\pi^2}{t}$) between short time and long time which mixes the scalar and spinor heat kernels.

\section{Borel transforms of short and long time expansions}
\label{sec:borel}

The two dimensional hyperboloid $\mathbb H^2$ is a harmonic space \cite{ruse}, so the heat kernel is a function only of the elapsed time $t$ and of the geodesic distance $\rho$ between the initial and final points on $\mathbb H^2$. An exact integral representation of the heat kernel is \cite{mckean,davies,Camporesi:1990wm,grigoryan,maxim}
\begin{eqnarray}
K(t, \rho)=\sqrt{2} \, \frac{e^{-t/4}}{(4 \pi t)^{3/2}}\int_\rho^\infty
  ds\, \frac{s\, e^{-s^2/(4 t)}}{\sqrt{\cosh(s) - \cosh(\rho)}}
    \label{eq:h2-rhoa}
  \end{eqnarray}
Changing variable by writing $s=\sqrt{\rho^2+4\pi^2\, u}$, this can be expressed as a Borel integral, suitable for generating a short time expansion:
  \begin{eqnarray}
K(t, \rho)
=
 2\pi \frac{e^{-t/4-\frac{\rho^2}{4t}}}{(4 \pi t)^{3/2}}\int_0^\infty \hskip -5pt 
  \frac{du}{\sqrt{u}}\,e^{-\frac{\pi^2 u}{t}} \hskip -5pt 
  \sqrt{ \frac{\pi^2\, u}{\sinh\left(\frac{\sqrt{\rho^2+4\pi^2 u}+\rho}{2}\right) \sinh\left(\frac{\sqrt{\rho^2+4\pi^2 u}-\rho}{2}\right)}}
  \label{eq:h2-rho}
  \end{eqnarray}

We therefore define the short time Borel transform function as:
\begin{eqnarray}
B_{\rm short}(u, \rho):= 
 \sqrt{ \frac{\pi^2\, u}{\sinh\left(\frac{\sqrt{\rho^2+4\pi^2 u}+\rho}{2}\right) \sinh\left(\frac{\sqrt{\rho^2+4\pi^2 u}-\rho}{2}\right)}}
\label{eq:borelhs}
\end{eqnarray}
\begin{figure}[htb]
\centering{\includegraphics[scale=0.75]{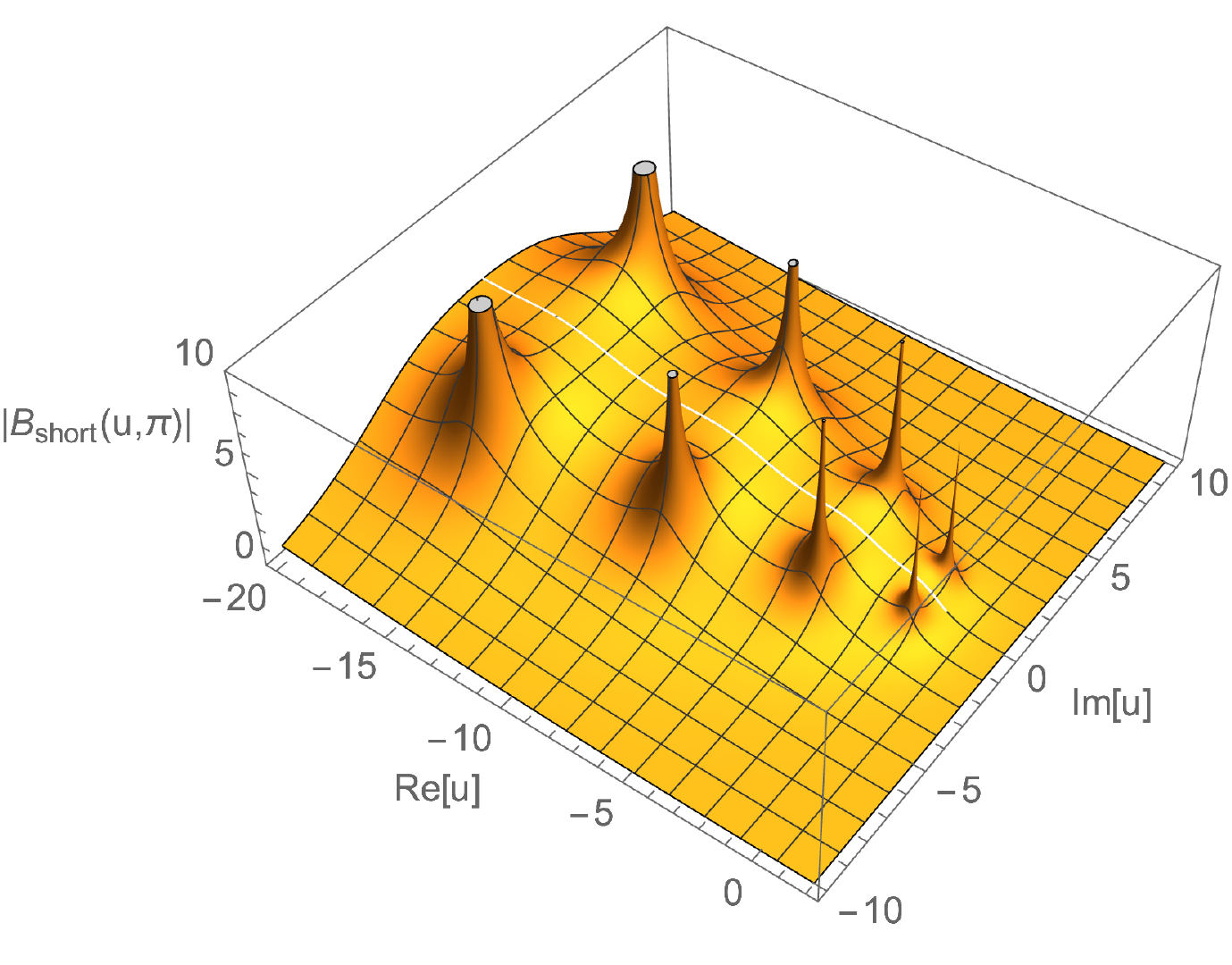}}
\caption{Plot of the absolute value of the short time Borel transform  $B_{\rm short}(u, \rho)$ in (\ref{eq:borelhs}), for  parameter choice $\rho=\pi$. Note that the Borel singularities have a  parabolic structure centered on the negative $u$ axis, as in (\ref{eq:borel-small-poles}). Contrast with Figure \ref{fig:borel-large-poles} for the Borel plane of the long time expansion.}
\label{fig:small-t-borel}
\end{figure}
This function has square root branch point singularities forming a parabola centered on the negative $u$ axis [see Figure \ref{fig:small-t-borel}]: 
\begin{eqnarray}
u_k^\pm=- k^2\pm i\,  k\, \rho/\pi \qquad, \quad k= 1, 2, 3, \dots
\label{eq:borel-small-poles}
\end{eqnarray}

\begin{figure}[htb]
\centering{\includegraphics[scale=0.7]{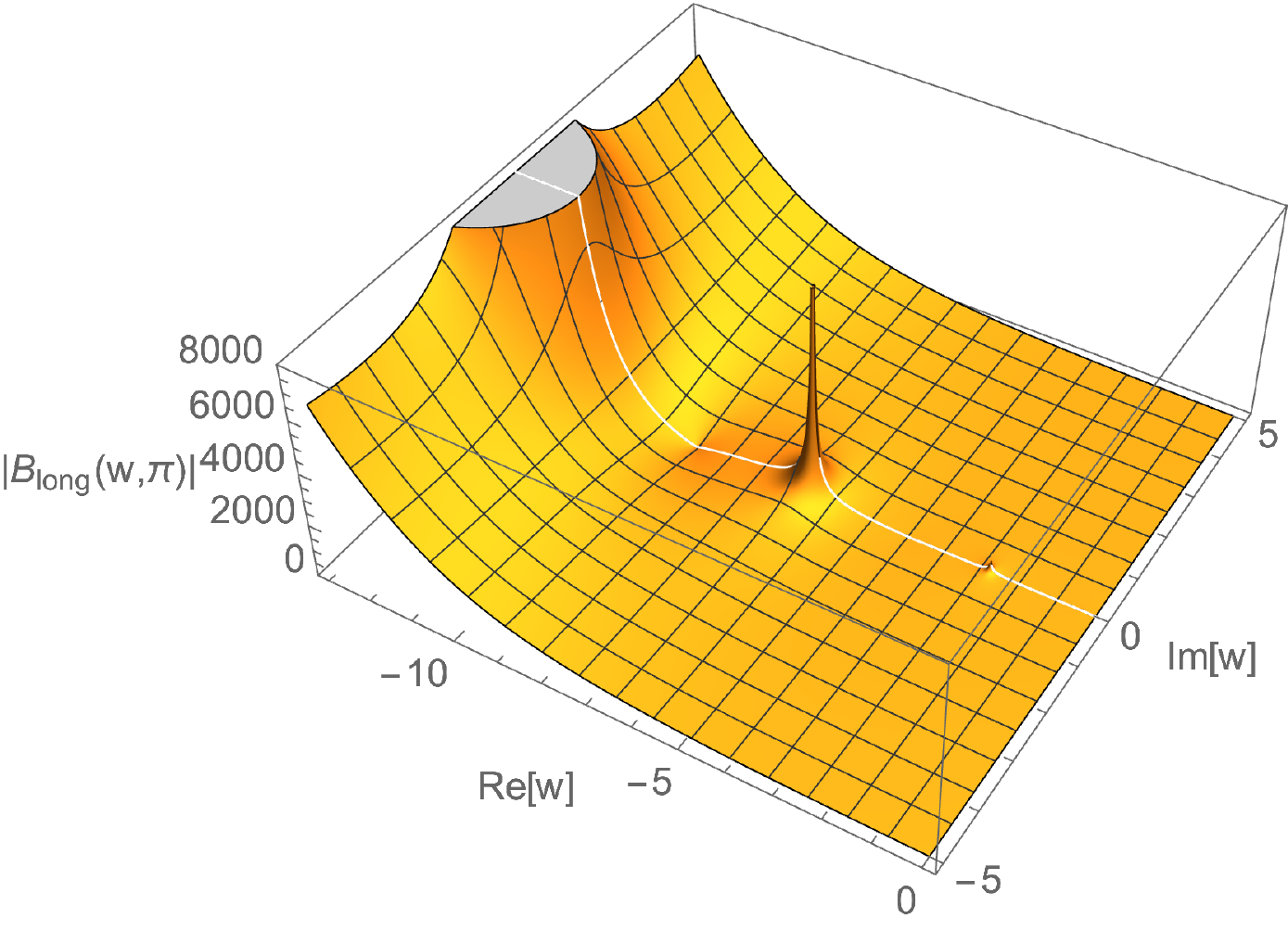}}
\caption{Plot of the absolute value of the long time Borel transform $B_{\rm long}(w, \rho)$ in (\ref{eq:borel-long}), for the parameter choice $\rho=\pi$. Note that the Borel singularities lie on the negative $w$ axis, as in (\ref{eq:borel-large-poles}). Contrast with Figure \ref{fig:small-t-borel} for the Borel plane of the short time expansion.}
\label{fig:borel-large-poles}
\end{figure}
An alternative integral representation of the heat kernel \cite{Camporesi:1990wm,Buchbinder:2014nia}, more suited for a large $t$ expansion reads:
\begin{eqnarray}
K(t, \rho)&=& \frac{e^{-t/4}}{2 \pi }\int_0^\infty
  dv\, e^{-v^2 t}\,v\, \tanh(\pi \,v)\, P_{-\frac{1}{2}+i\, v}(\cosh(\rho))
  \label{eq:h2-rho-large-v} \\
  &=& \frac{e^{-t/4}}{4 \pi }\int_0^\infty
  dw\, e^{-w\,  t} \, \tanh(\pi \sqrt{w})\, P_{-\frac{1}{2}+i\sqrt{w}}(\cosh(\rho))
    \label{eq:h2-rho-large}
  \end{eqnarray}
  Here $P_{-\frac{1}{2}+i\, v}$ is the Legendre function of the first kind. The second expression  has the form of a Borel integral, but for an expansion in inverse powers of $t$. We define the long time Borel transform function:
  \begin{eqnarray}
  B_{\rm long}(w, \rho):= \tanh(\pi \sqrt{w})\, P_{-\frac{1}{2}+i\sqrt{w}}(\cosh(\rho))
  \label{eq:borel-long}
  \end{eqnarray}
  The singularities of  $B_{\rm long}(w, \rho)$ lie along the negative real $w$ axis, corresponding to the singularities  of the $\tanh(\pi \sqrt{w})$ factor in (\ref{eq:borel-long}) [see Figure \ref{fig:borel-large-poles}]:
\begin{eqnarray}
w_k=-\left(k+\frac{1}{2}\right)^2 \quad, \quad k=0, 1, 2, 3, ...
\label{eq:borel-large-poles}
\end{eqnarray}
   Notice that the Borel plane structure of the  small $t$ expansion is very different from the  Borel plane structure of the  large $t$ expansion.

\subsection{Borel structure and large-order growth for the short time expansion}
\label{sec:borel-small-t}

The small $t$ expansion of the heat kernel $K(t, \rho)$ can be obtained by expanding the short time Borel transform function $B_{\rm short}(u, \rho)$ in (\ref{eq:borelhs}) at small $u$:
\begin{eqnarray}
B_{\rm short}(u, \rho)&=&
\sqrt{\frac{\rho}{\sinh(\rho)}}  \left[1+\frac{\pi ^2 u (1-\rho  \coth (\rho))}{2 \rho ^2} 
 +\frac{\pi ^4 u^2 \left(-15-8 \rho ^2+9 \rho ^2 \coth ^2(\rho )+6 \rho  \coth (\rho )\right)}{24 \rho ^4}+\dots\right]
\nonumber\\
   &:=&
\sqrt{\frac{\rho}{\sinh(\rho)}} \sum_{n=0}^\infty c_n(\rho, {\rm coth}(\rho))\, u^{n}
\label{eq:u-exp}
\end{eqnarray}
The coefficients $c_n$ are simple polynomials in $1/\rho$ and ${\rm coth}(\rho)$, whose explicit expressions are cumbersome and not essential here (we normalize so that $c_0=1$). 
This yields the  small $t$ asymptotic expansion for (\ref{eq:h2-rho}):
\begin{eqnarray}
K(t, \rho)\sim \sqrt{\frac{\rho}{\sinh(\rho)}} \, \frac{e^{-t/4-\rho^2/(4t)}}{4 \pi^{3/2}\, t} \sum_{n=0}^\infty c_n(\rho, {\rm coth}(\rho))\,\Gamma\left(n+\frac{1}{2}\right) \left(\frac{t}{\pi^2}\right)^{n}
\, , \, t\to 0^+
\label{eq:small-t-asymptotic}
\end{eqnarray}
The {\it  leading} short time behavior is
  \begin{eqnarray}
K(t, \rho)\sim  \frac{e^{-\rho^2/(4t)}}{4 \pi t} \sqrt{\frac{\rho}{\sinh(\rho)}} e^{-t/4}+\dots 
 \qquad, \quad t\to 0^+
 \label{eq:short-leading}
\end{eqnarray}
the first factor of which is the typical form of the short time asymptotics for heat propagation on a 2 dimensional manifold. 
\begin{figure}[htb]
\centering{\includegraphics[scale=0.5]{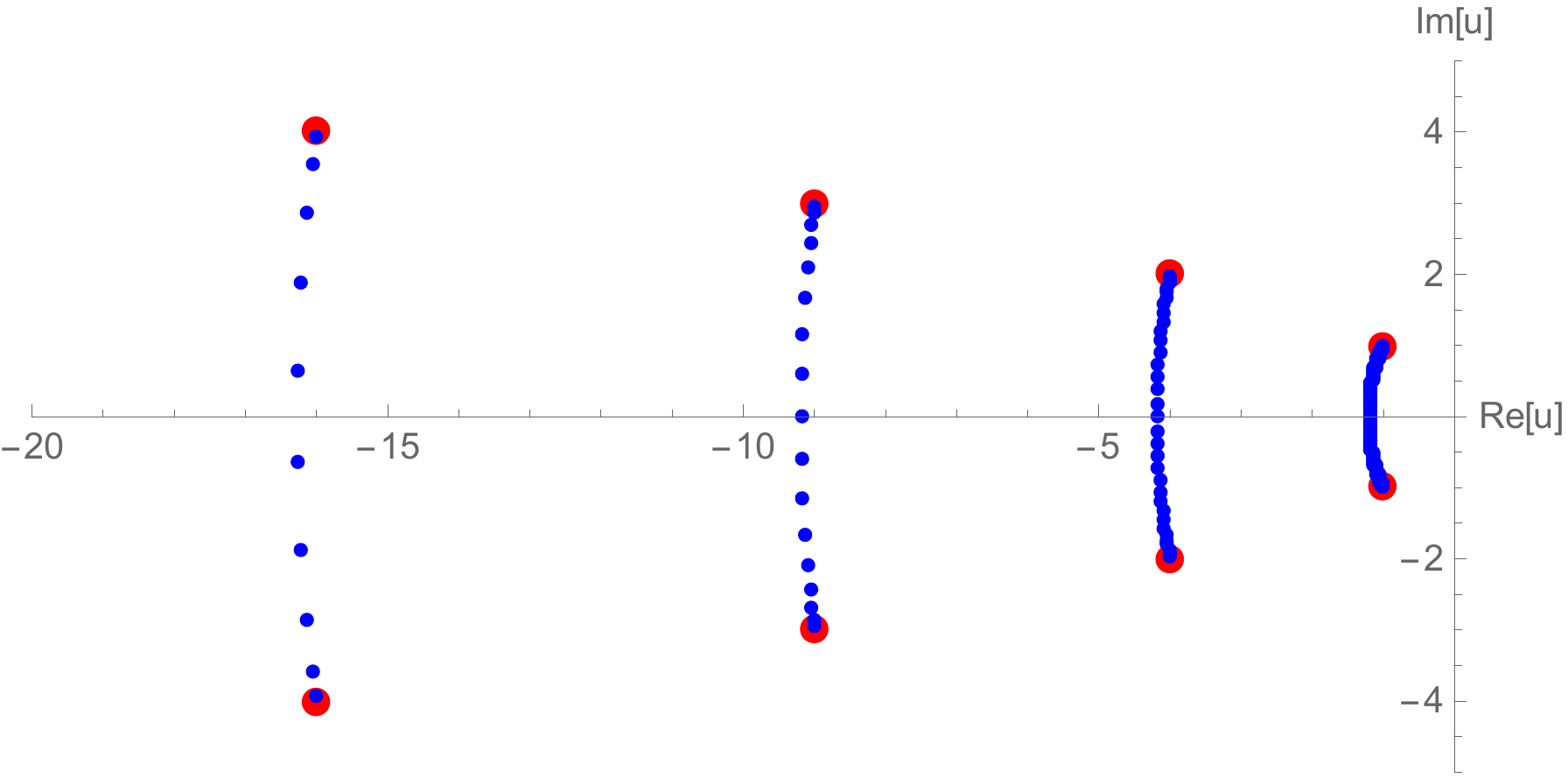}}
\caption{Poles (blue points) of a high-order diagonal Pad\'e approximant of the short time Borel transform $B_{\rm short}(u, \rho)$ in (\ref{eq:borelhs}), plotted here for $\rho=\pi$. The actual Borel singularities (\ref{eq:borel-small-poles}) are shown as red dots. Since these are branch points, Pad\'e represents them as the accumulation points of arcs of poles. Compare with  Figure \ref{fig:small-t-borel}. }
\label{fig:small-t-poles}
\end{figure}

To probe the asymptotic nature of the short time expansion (\ref{eq:small-t-asymptotic}) it is instructive to construct a Pad\'e approximant of a truncation of the Borel expansion (\ref{eq:u-exp}). Figure \ref{fig:small-t-poles} shows the Pad\'e poles for a diagonal Pad\'e approximant of the small $t$ Borel function $B_{\rm short}(u, \rho)$, for $\rho=\pi$. We see the parabola of Borel singularities, represented by Pad\'e as the accumulation points of arcs of poles. This reflects the fact that the Borel singularities are branch points rather than poles, and Pad\'e places arcs of poles representing the cuts of the associated minimal capacitor in the electrostatic interpretation of Pad\'e analysis \cite{Stahl,Saff,Costin:2020pcj,Costin:2021bay}. 
\begin{figure}[htb]
\centering{\includegraphics[scale=0.7]{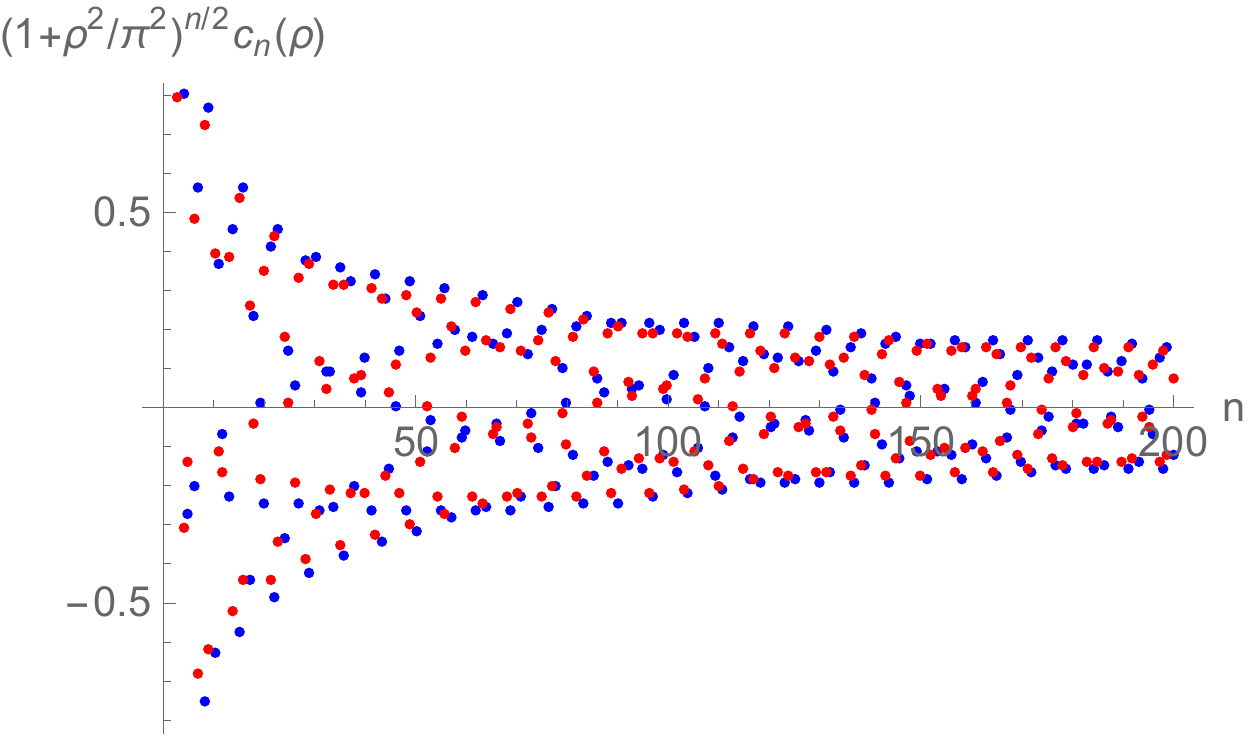}}\\
\centering{\includegraphics[scale=0.7]{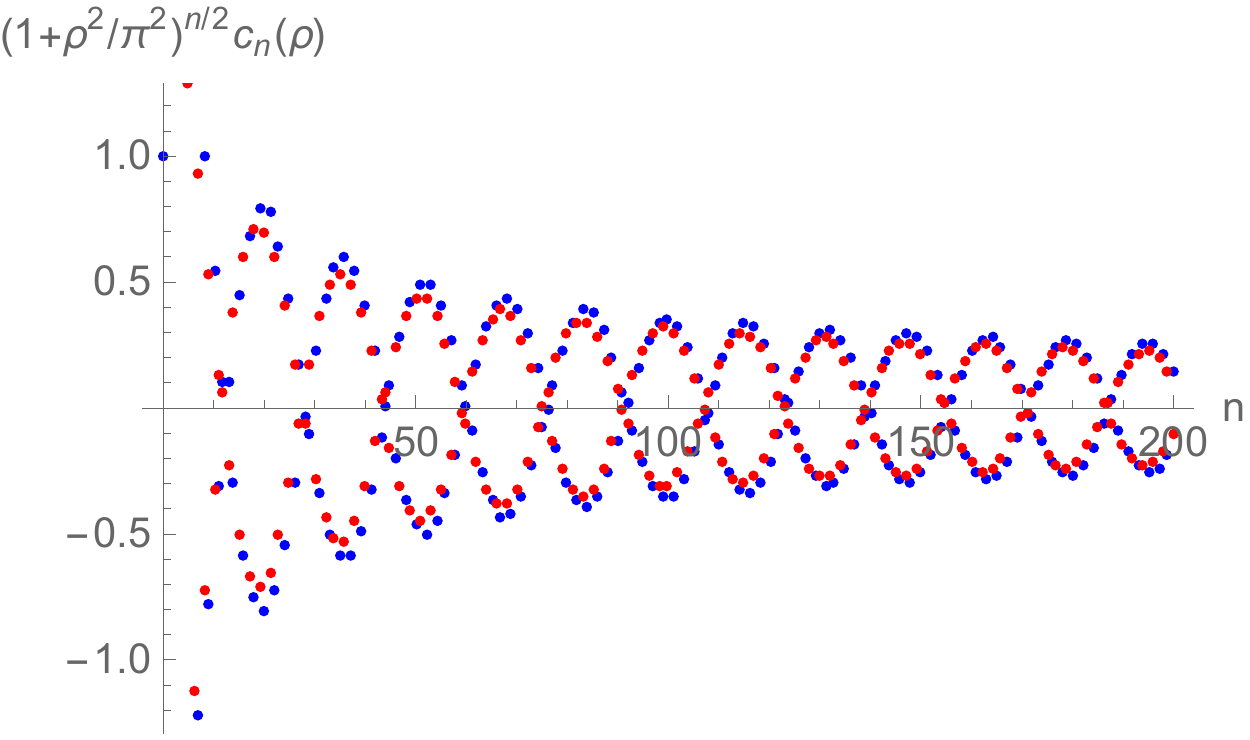}}
\caption{These plots show the large order ($n\to\infty$) behaviour of the expansion coefficients $c_n(\rho, \coth(\rho))$ in the expansion (\ref{eq:u-exp}) of the short time Borel transform function. The coefficients have been multiplied by $\left(1+\frac{\rho^2}{\pi^2}\right)^{n/2}$, and plotted for two different values of $\rho$: $\rho=\frac{\pi}{2}$, and $\rho=\frac{\pi}{5}$. The blue dots are the exact values, while the red dots are the asymptotic estimate in (\ref{eq:cn-asym}). }
\label{fig:large-order-cn}
\end{figure}

For a given $\rho$, the 
coefficients of the short time expansion (\ref{eq:small-t-asymptotic}) grow factorially fast in magnitude due to the 
$\Gamma\left(n+\frac{1}{2}\right)$ factor. The $c_n(\rho, \coth(\rho))$ coefficient factors in  (\ref{eq:small-t-asymptotic})  have an intricate oscillatory sign pattern which depends sensitively on the magnitude of $\rho$. This oscillatory behavior is due to interference between the complex Borel singularities. The leading large order contribution to this oscillatory behavior can be understood from the product representation of the sinh functions in $B_{\rm short}(u, \rho)$:
\begin{eqnarray}
B_{\rm short}(u, \rho) &=& \frac{\sqrt{\frac{\rho}{\sinh(\rho)}}}{\sqrt{\prod_{k=1}^\infty \left(1+\frac{2u}{k^2\left(1+\frac{\rho^2}{k^2\, \pi^2}\right)} +\frac{u^2}{k^4\left(1+\frac{\rho^2}{k^2\, \pi^2}\right)}\right)}}
\label{eq:bshort-product}
\end{eqnarray}
The dominant large order growth of the coefficients $c_n(\rho, \coth(\rho))$ comes from the leading ($k=1$) singularity, which gives the leading  estimate
\begin{eqnarray}
c_n(\rho, \coth(\rho))\sim -\sqrt{\pi} (-1)^n \left(1+\frac{\rho^2}{\pi^2}\right)^{-n/2} P_n\left(\frac{1}{\sqrt{1+\frac{\rho^2}{\pi^2}}}\right)
\quad, \quad n\to \infty
\label{eq:cn-asym}
\end{eqnarray}
where $P_n$ is the Legendre polynomial. The overall normalization coefficient, related to  the cumulative effect of the other singularities, has been fitted. The resulting oscillatory behavior is illustrated in Figure \ref{fig:large-order-cn}, for two different values of $\rho$: $\rho=\frac{\pi}{2}$ and for $\rho=\frac{\pi}{5}$. Interference with more distant singularities produces further fine structure of the large order behavior, but the leading pattern is captured by the simple formula (\ref{eq:cn-asym}).

\subsection{Borel structure and large-order growth for the long time expansion}
\label{sec:borel-large-t}

The large $t$ expansion of $K(t, \rho)$ may be generated from the small $w$ expansion of the long time Borel 
transform (\ref{eq:borel-long}):
\begin{eqnarray}
B_{\rm long}(w, \rho)
&:=& \frac{1}{2\sqrt{\pi}}\sum_{n=0}^\infty d_n(\rho) w^{n+\frac{1}{2}}
\label{eq:borelhl-exp}
\end{eqnarray}
The coefficients $d_n(\rho)$  are expressed in terms of the elliptic integral of the first kind, $\mathbb K$, and derivatives of the Legendre function with respect to its index. This produces a long time asymptotic expansion of the heat kernel: 
  \begin{eqnarray}
K(t, \rho)\sim \frac{e^{-t/4}}{(4 \pi t)^{3/2}}  \sum_{n=0}^\infty \frac{d_n(\rho)}{t^{n}}
\qquad, \quad t\to +\infty
\label{eq:k-long-t}
\end{eqnarray}
The {\it leading} large $t$ term is
  \begin{eqnarray}
K(t, \rho)\sim  2\pi\,  \mathbb K\left(-\sinh^2\left(\frac{\rho}{2}\right)\right) \frac{e^{-t/4}}{(4 \pi t)^{3/2}} 
+\dots \qquad, \quad t\to +\infty
\label{eq:klong-leading}
\end{eqnarray}
where $\mathbb K$ is the elliptic integral of the first kind. 
An alternative derivation of the large $t$ expansion can be generated by expanding the  exponential factor in (\ref{eq:h2-rhoa}) in powers of $\frac{s^2}{4t}$ (instead of expanding the $1/\sqrt{\cosh(s)-\cosh(\rho)}$ factor), and then evaluating the resulting $s$ integrals. 

Analysis of the long time expansion (\ref{eq:k-long-t}), also as a function of $\rho$, reveals it to be asymptotic, with factorially divergent (and sign alternating) expansion coefficients $d_n(\rho)$. The leading large order growth is independent of the geodesic distance $\rho$, while the subleading corrections depend on $\rho$ [see also Eqs (\ref{eq:dn-zero-a})-(\ref{eq:dn-zero-b}) for the closed-form expression when $\rho=0$] :
\begin{eqnarray}
d_n(\rho) \sim \frac{16}{\sqrt{\pi}} (-4)^n \Gamma\left(n+\frac{3}{2}\right) \left(1+\frac{\cosh(\rho)}{9^{n+1}}+\dots\right) \qquad, \quad n\to\infty
\label{eq:dn-large}
\end{eqnarray}
Correspondingly, one may construct a Pad\'e approximant of the long time Borel transform $B_{\rm long}(w, \rho)$, whose singularities lie along the negative $w$ axis, coinciding with the analytic result in (\ref{eq:borel-large-poles}).

\subsection{Expansions at Intermediate Time Scales}
\label{sec:borel-all-t}

The previous expansions have assumed that the time $t$ is either the smallest or the largest asymptotic parameter. However, the time $t$ could also be measured relative to the geodesic distance $\rho$ (in units in which the radius of curvature and the diffusion constant have been normalized to 1). This means that there are different asymptotic regions.\cite{davies,anker}.
In general we can expand the heat kernel as follows. From the Legendre expansion of (\ref{eq:h2-rhoa}) we obtain
\begin{eqnarray}
K(t, \rho)&=& 2 \, \frac{e^{-\frac{t}{4}}}{(4 \pi t)^{3/2}} \sum_{k=0}^\infty P_k(\cosh(\rho)) \int_\rho^\infty
  ds\, s\, e^{-\frac{s^2}{4 t}-\left(k+\frac{1}{2}\right)s}
    \label{eq:h2-rho-all-1}
 \\
    &=& 4t \frac{e^{-\frac{t}{4}-\frac{\rho^2}{4t}}}{(4 \pi t)^{3/2}}  \sum_{k=0}^\infty e^{-\left(k+\frac{1}{2}\right)\rho}  P_k(\cosh(\rho)) 
   \left[\frac{\rho}{\rho+(2k+1)t}+\frac{\sqrt{t}}{2} \left(k+\frac{1}{2}\right)  e^{\frac{(t+2k t+\rho)^2}{4t}} \Gamma\left(-\frac{1}{2}, \frac{(t+2k t+\rho)^2}{4t}\right) \right]
    \label{eq:h2-rho-all-2}
  \end{eqnarray}
  This explains why there are different regions of the asymptotics, depending on the relative size of $t$ and $\rho$, with appropriate dimensional scalings. We will see below that these expansions in terms of incomplete gamma functions are surprisingly accurate.

\section{Diagonal heat kernel asymptotics}
\label{sec:diagonal}

More analytic detail is available for the {\it diagonal} heat kernel, $K(t, 0)$, for heat propagation on $\mathbb H^2$ where the final point coincides with the initial point ({\it i.e.}, $\rho=0$).

\subsection{Diagonal heat kernel asymptotics at short time $t$}
\label{sec:diagonal-small-t}

When $\rho=0$, the short time expression (\ref{eq:h2-rho}) simplifies to
\begin{eqnarray}
K(t, 0)
 & = &\frac{e^{-t/4}}{(4 \pi t)^{3/2}}\int_0^\infty
  ds\, \frac{s\, e^{-s^2/(4 t)}}{\sinh(s/2)}
    \label{eq:h2-rho-zero}
  \end{eqnarray}
  In the $\rho\to 0$ limit the complex conjugate pairs of branch point Borel singularities in (\ref{eq:borel-small-poles}) merge to become poles on the negative real axis  (see also Figures \ref{fig:small-t-borel} and \ref{fig:small-t-poles}).

 The asymptotic small $t$ expansion of $K(t, 0)$ follows from the expansion:
 \begin{eqnarray}
  \frac{s}{\sinh(s/2)}= 2\sum_{n=0}^\infty s^{2n} \frac{B_{2n}\left(\frac{1}{2}\right)}{(2n)!} 
  \label{eq:sinh-exp}
  \end{eqnarray}
  where $B_{2n}$ is the Bernoulli polynomial, and we note that $B_{2n}\left(\frac{1}{2}\right)$ can be written as
  \begin{eqnarray}
  B_{2n}\left(\frac{1}{2}\right)=(-1)^n \frac{2(2n)!}{(2\pi)^{2n}} \eta(2n)
  \label{eq:bernoulli}
  \end{eqnarray}
in terms of the Dirichlet eta function  
\begin{eqnarray}
\eta(2n)=\sum_{k=1}^\infty \frac{(-1)^{k+1}}{k^{2n}}
\label{eq:dirichlet}
\end{eqnarray}
  Thus we obtain a simple closed-form expression for the short time asymptotic expansion of the diagonal heat kernel:
\begin{eqnarray}
K(t, 0)\sim \frac{e^{-t/4}}{2 \pi^{3/2} t}  \sum _{n=0}^\infty (-1)^n  \eta (2 n)\, \Gamma \left(n+\frac{1}{2}\right)\, \left(\frac{t}{\pi^2}\right)^{n}
\quad, \quad t\to 0^+
  \label{eq:h2-rho-zero-smallt}
  \end{eqnarray}  
  The {\it leading} short time term is 
  \begin{eqnarray}
K_{\rm leading} (t, 0)\sim \frac{e^{-t/4}}{4 \pi\, t}  \quad, \quad t\to 0^+
  \label{eq:h2-rho-zero-smallt-leading}
  \end{eqnarray}  
  which is consistent with the $\rho\to 0$ limit of (\ref{eq:short-leading}).
  
\begin{figure}[htb]
 \centerline{\includegraphics[scale=.7]{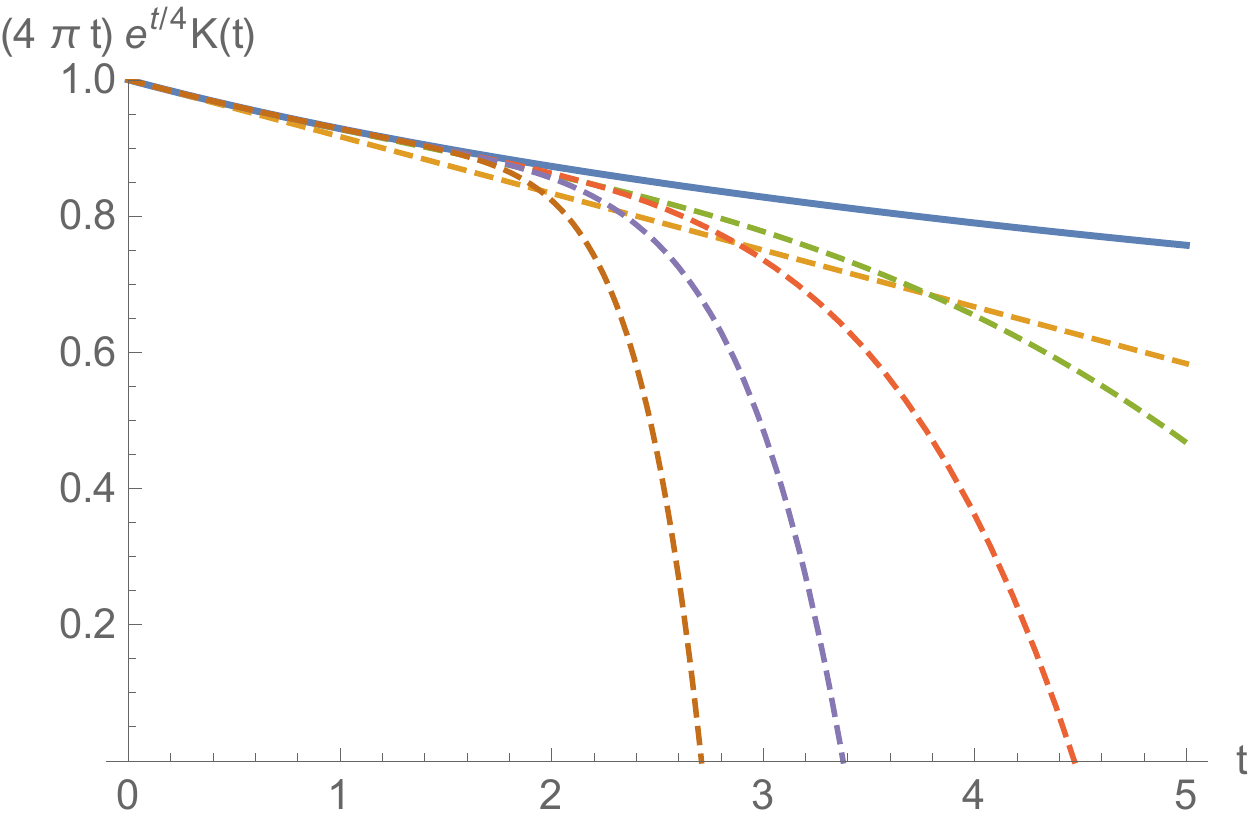}}
 \caption{The solid blue curve is the exact diagonal heat kernel $K(t, 0)$  in (\ref{eq:h2-rho-zero}), normalized by the leading short time expression $K_{\rm leading}(t, 0)=e^{-t/4}/(4\pi t)$. The dashed curves show the same ratio, using successively $1, 3, 5, 7, 9$ terms of the asymptotic expansion in (\ref{eq:h2-rho-zero-smallt}).}
 \label{fig:kshort-expansions}
 \end{figure}

 The expansion coefficients in (\ref{eq:h2-rho-zero-smallt}) grow factorially fast in magnitude, and alternate in sign.  Recall that $\eta(2n)$ approaches $1$ exponentially quickly. Figure \ref{fig:kshort-expansions} illustrates the typical behaviour of truncations of an alternating sign factorially divergent asymptotic series.
 Note also that this expansion agrees with the $\rho\to 0$ limit of the general short time asymptotic expansion in (\ref{eq:small-t-asymptotic}), based on the $\rho\to 0$ limit of the Borel expansion in (\ref{eq:u-exp}), because the Borel coefficients $c_n(\rho, \coth(\rho))\to 2 (-1)^n \eta(2n)$ as $\rho\to 0$.

 A resummed form of this short time asymptotic expansion  (\ref{eq:h2-rho-zero-smallt}) can be obtained using the instanton sum form of the Dirichlet eta function in (\ref{eq:dirichlet}), combined with elementary properties of the incomplete gamma function, which has an asymptotic expansion:
    \begin{eqnarray}
  x^{-\alpha}\, e^x\, \Gamma(1+\alpha, x)\sim \frac{1}{\Gamma(-\alpha)} \sum_{n=0}^\infty (-1)^n \Gamma(n-\alpha)\frac{1}{x^{n}}\quad, \quad x\to +\infty
  \label{eq:gamma}
  \end{eqnarray}  
This leads to a ``gamma-resummed''  form of the short time expansion
   \begin{eqnarray}
K_{\rm gamma}(t, 0)\sim
\frac{e^{-t/4}}{4 \pi t}  \left(1+\frac{\pi}{\sqrt{t}}\sum_{k=1}^\infty (-1)^k k\, e^{\pi^2k^2/t}\,\Gamma\left(-\frac{1}{2}, \frac{\pi^2 k^2}{t} \right) \right)
\,, \, t\to 0^+
  \label{eq:h2-rho-zero-smallt-gamma}
  \end{eqnarray}  
The first term is the leading short time behaviour from (\ref{eq:short-leading}), and the corrections are expressed as an infinite sum in terms of incomplete gamma functions.
 An analogous resummed expression is obtained below for the long time expansion -- see (\ref{eq:h2-rho-zero-larget-gamma}).
Note that the short-time gamma-resummed expansion (\ref{eq:h2-rho-zero-smallt-gamma}) can also be obtained from the $\rho=0$ limit of the long time expression (\ref{eq:h2-rho-large-v}),
 by expanding the $\tanh(\pi v)$ factor as
\begin{eqnarray}
\tanh(\pi\, v)=1+2\sum_{k=1}^\infty(-1)^k e^{-2\pi k v}
\label{eq:tanh}
\end{eqnarray}
The resulting $v$ integrals lead again to (\ref{eq:h2-rho-zero-smallt-gamma}).
     \begin{figure}[htb]
 \centerline{\includegraphics[scale=.7]{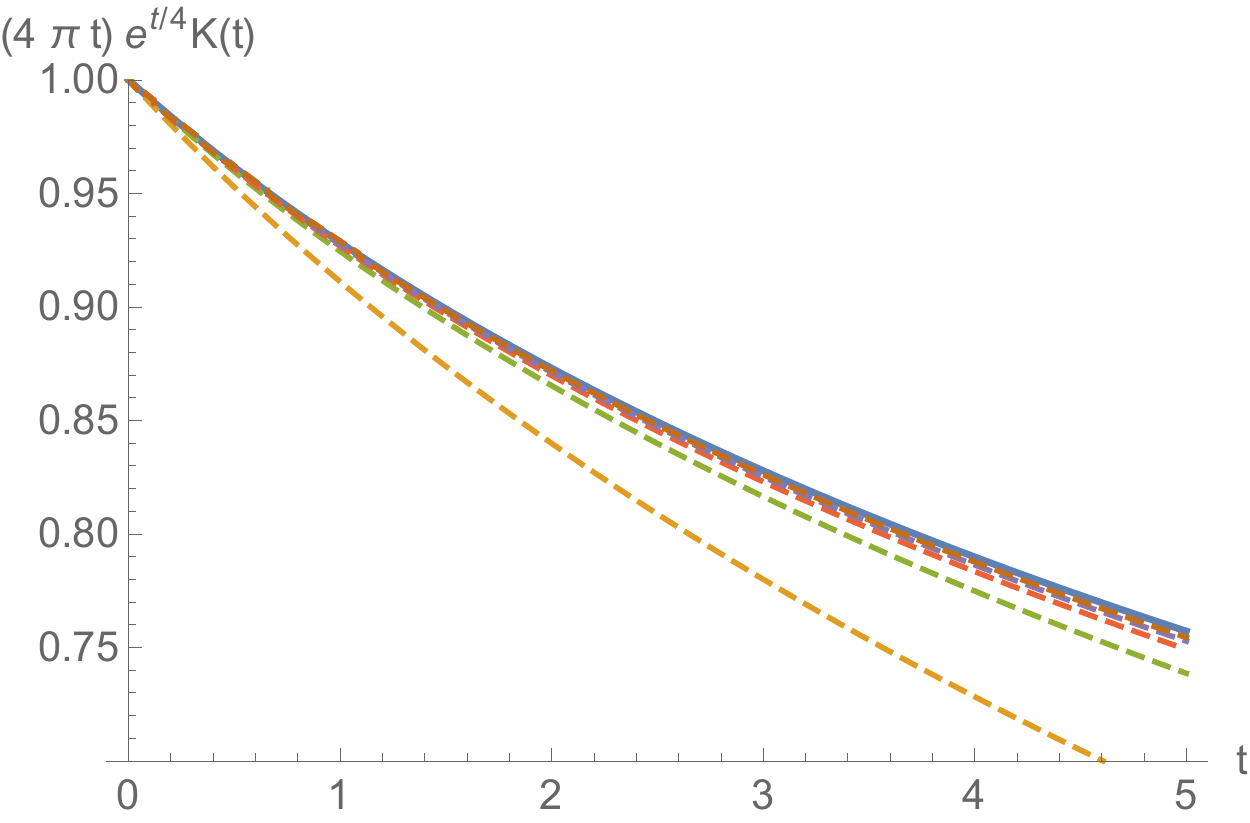}}
 \caption{The solid blue curve is the exact diagonal heat kernel $K(t, 0)$  in (\ref{eq:h2-rho-zero}), normalized by the leading short time expression $K_{\rm leading}(t, 0)=e^{-t/4}/(4\pi t)$. The dashed curves show the same ratio, using successively $1, 3, 5, 7, 9$ terms of the convergent gamma-resummed expansion in (\ref{eq:h2-rho-zero-smallt-gamma}). Note the significant improvement over the corresponding asymptotic expansions plotted in Figure \ref{fig:kshort-expansions}, which has the same colour scheme with respect to the number of terms in each expansion.}
 \label{fig:kshort-gamma}
 \end{figure}

In contrast to the short-time asymptotic expansion (\ref{eq:h2-rho-zero-smallt}), the short-time gamma-resummed expansion (\ref{eq:h2-rho-zero-smallt-gamma}) is convergent.  Figure \ref{fig:kshort-gamma} shows the dramatic improvement of this resummed form of the short time expansion when it is extrapolated from short time to long time.
 Moreover, the analytic properties of the incomplete gamma functions encode the correct analytic continuation properties of the diagonal heat kernel. This will be important in Sections  \ref{sec:schr} and \ref{sec:sphere} below, where we discuss the analytic continuation of $K(t, 0)$ under $t\to i t$ and $t\to -t$, which are physically relevant for the analytic continuation of the heat kernel to the Schr\"odinger kernel (Section \ref{sec:schr}), and of the heat kernel on the hyperboloid to the heat kernel on the sphere (Section \ref{sec:sphere}), respectively.

\subsection{Diagonal heat kernel asymptotics at long time  $t$}
\label{sec:diagonal-large-t}
\begin{figure}[htb]
 \centerline{\includegraphics[scale=.7]{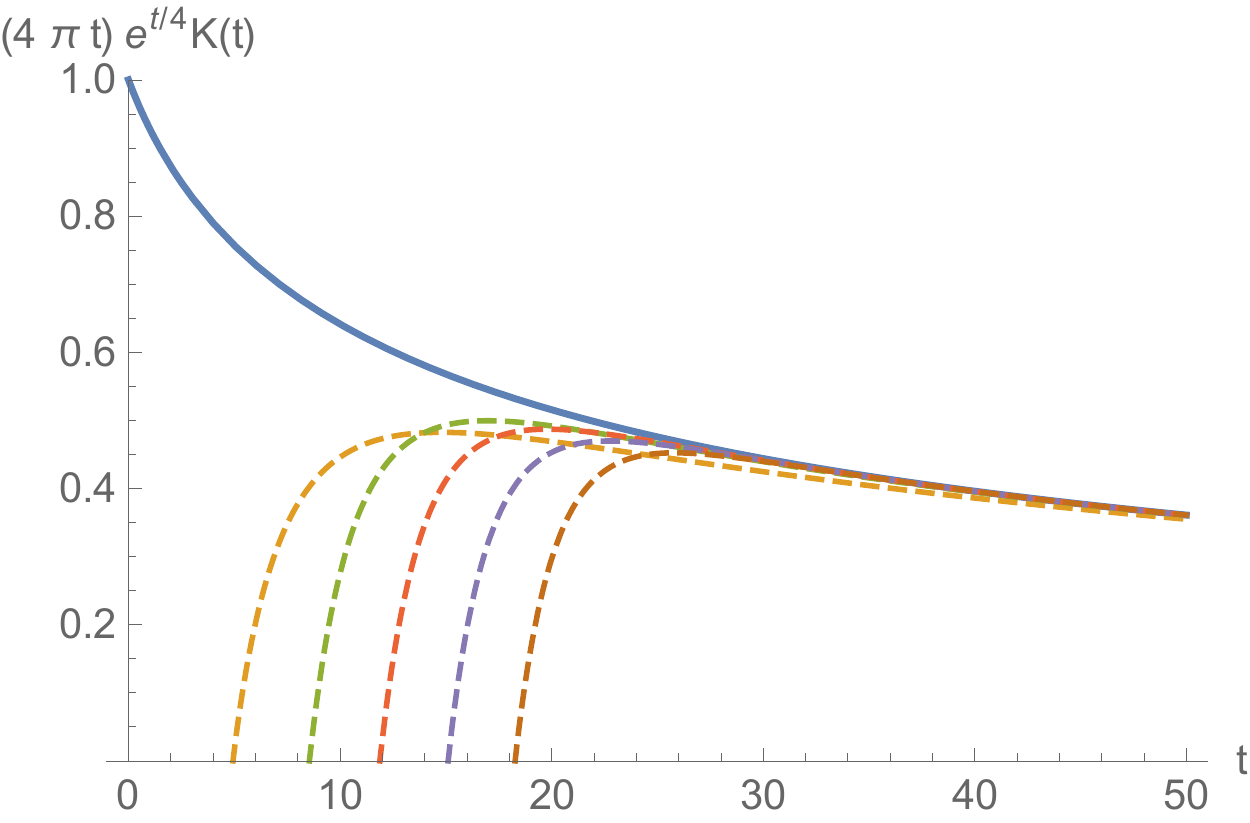}}
 \caption{The solid blue curve is the exact long time diagonal heat kernel $K(t, 0)$  in (\ref{eq:h2-rho-large-d}), normalized by the leading short time expression $K_{\rm leading}(t, 0)=e^{-t/4}/(4\pi t)$. The dashed curves show the same ratio, using successively $1, 3, 5, 7, 9$ terms of the long time asymptotic expansion in (\ref{eq:h2-rho-zero-larget}). Compare with the corresponding short time expansions in Figure \ref{fig:kshort-expansions}.}
 \label{fig:klarge-expansions}
 \end{figure}

When $\rho=0$, the long time heat kernel expression (\ref{eq:h2-rho-large}) simplifies to
 \begin{eqnarray}
K(t, 0)&=& \frac{e^{-t/4}}{2 \pi }\int_0^\infty
  dv\,v\, \tanh(\pi \,v)\,  e^{-v^2 t}
    \label{eq:h2-rho-large-d}
  \end{eqnarray} 
  The long time asymptotics follows from the small $v$ expansion:
  \begin{eqnarray}
  v\, \tanh(\pi \,v) = -\frac{2}{\pi} \sum_{n=0}^\infty (-1)^n (2^{2n}-1) \zeta(2n)\, v^{2n} 
  \label{eq:tanh2}
  \end{eqnarray}
This leads to an alternating-sign factorially divergent asymptotic expansion as $t\to +\infty$ (see Figure \ref{fig:klarge-expansions}):
  \begin{eqnarray}
K(t, 0)
\sim  \frac{16}{\sqrt{\pi }}  \, \frac{e^{-t/4}}{(4 \pi t)^{3/2}}  \sum _{n=0}^\infty (-1)^n \left(1-2^{-(2n+2)} \right) \zeta (2 n+2) \Gamma \left(n+\frac{3}{2}\right)\, \left(\frac{4}{t}\right)^n
  \label{eq:h2-rho-zero-larget}
  \end{eqnarray}  
  We can alternatively generate this large $t$ expansion by expanding the Gaussian factor $e^{-s^2/(4 t)}$ in the short time diagonal heat kernel (\ref{eq:h2-rho-zero}), using the integral identity
  \begin{eqnarray}
  \int_0^\infty ds\, \frac{s^{2n+1}}{\sinh(s/2)} =2\left(2^{2n+2}-1\right)\zeta(2n+2)\,\Gamma(2n+2)
  \label{eq:sinh-integral}
  \end{eqnarray}
  
  In the notation of (\ref{eq:k-long-t}) the expression (\ref{eq:h2-rho-zero-larget}) yields an exact closed-form expression for the long time expansion coefficients $d_n(\rho)$ when $\rho=0$:
  \begin{eqnarray}
  d_n(0)&=&  \frac{16}{\sqrt{\pi }}  (-4)^n \left(1-2^{-(2n+2)} \right) \zeta (2 n+2) \Gamma \left(n+\frac{3}{2}\right)
  \label{eq:dn-zero-a} \\
  &\sim& \frac{16}{\sqrt{\pi}} (-4)^n \left(1+\frac{1}{9^{n+1}}+\dots\right) \Gamma \left(n+\frac{3}{2}\right)\quad, \quad n\to\infty
  \label{eq:dn-zero-b}
  \end{eqnarray}
Note that  the $1/2^{2n+2}$ part of $\zeta (2 n+2)$ exactly cancels with the contribution from the $\left(1-2^{-(2n+2)}\right)$ factor, leaving the leading large order correction as $1/3^{2n+2}=1/9^{n+1}$.
This large order growth result (\ref{eq:dn-zero-b}) is consistent with the $\rho\to 0$ limit of the observed large order behaviour at nonzero $\rho$ in (\ref{eq:dn-large}). Furthermore, we see that the leading long time behaviour
    \begin{eqnarray}
K(t, 0)\sim
\pi^2 \frac{e^{-t/4}}{(4\pi t)^{3/2}} 
\qquad, \quad t\to +\infty
  \label{eq:h2-rho-zero-larget-leading}
  \end{eqnarray}
 agrees with the $\rho\to 0$ limit of (\ref{eq:klong-leading}), as $\mathbb K(0)=\frac{\pi}{2}$.

        \begin{figure}[htb]
 \centerline{\includegraphics[scale=.7]{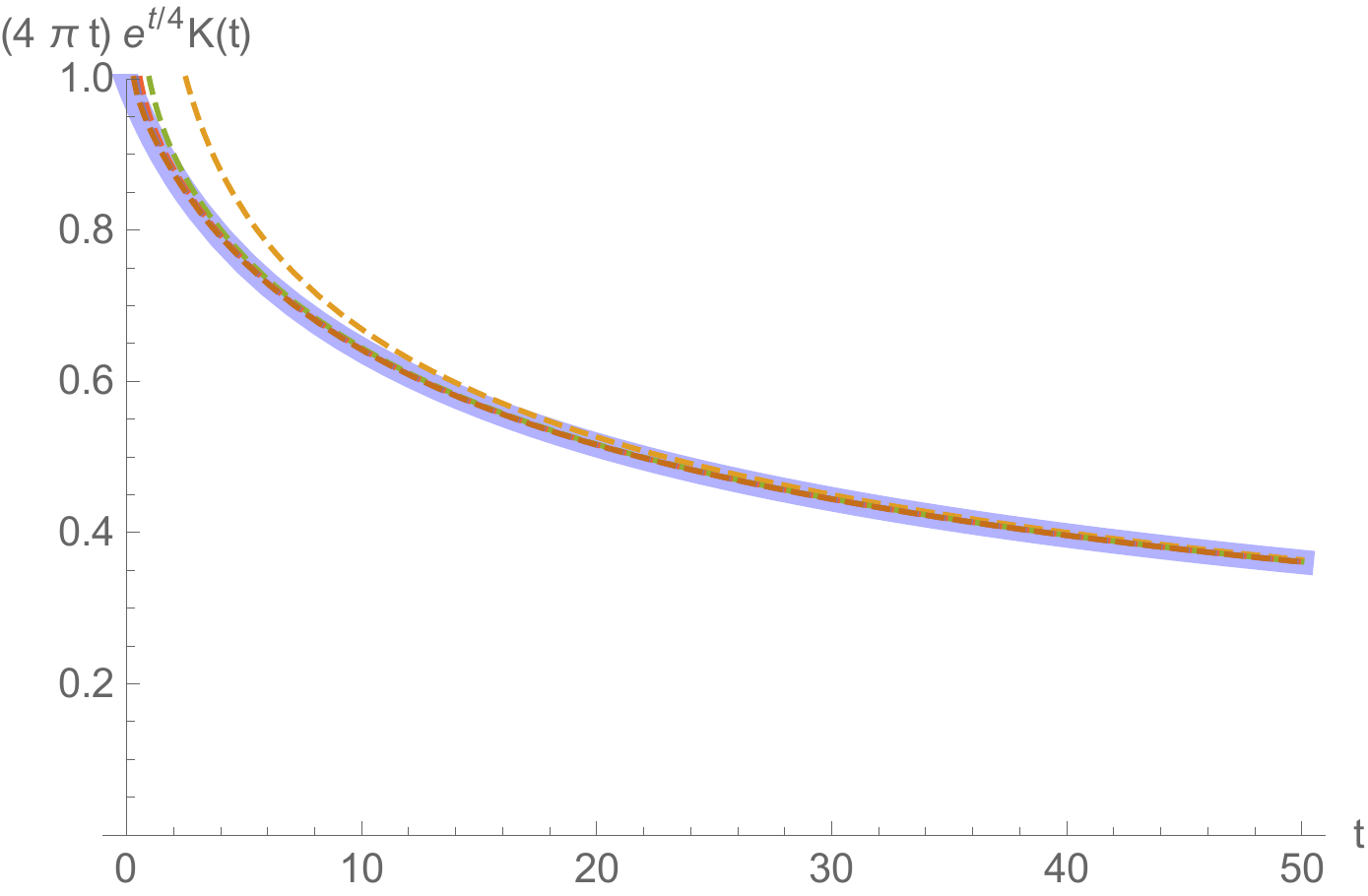}}
 \caption{The solid blue curve is the exact long time diagonal heat kernel $K(t, 0)$  in (\ref{eq:h2-rho-large-d}), normalized by the leading short time expression $K_{\rm leading}(t, 0)=e^{-t/4}/(4\pi t)$. The dashed curves show the same ratio, using successively $1, 3, 5, 7, 9$ terms of the long time gamma-resummed asymptotic expansion in (\ref{eq:h2-rho-zero-larget-gamma}). Compare with the corresponding short time expansions in Figure \ref{fig:kshort-gamma}.}
 \label{fig:klarge-gamma}
 \end{figure}
 
 We obtain a gamma-resummed expression in terms of incomplete gamma functions by  replacing the zeta function factor in (\ref{eq:h2-rho-zero-larget}) by an ``instanton sum'', $\zeta(2n+2)=\sum_{k=1}^\infty \frac{1}{k^{2n+2}}$, and using the incomplete gamma function asymptotics (\ref{eq:gamma}):
 \begin{eqnarray}
K_{\rm gamma}(t, 0)&\sim&
\frac{e^{-t/4}}{(4 \pi t)^{3/2} }  \left(\pi^2 
- 3t^{3/2} \sum_{k=1}^\infty k\left[\frac{1}{2} e^{k^2 t/4}\,\Gamma\left(-\frac{3}{2},\frac{ k^2 t}{4}\right)
- e^{k^2 t}\,\Gamma\left(-\frac{3}{2}, k^2 t \right)   \right]\right)
\quad, \quad t\to +\infty
  \label{eq:h2-rho-zero-larget-gamma}
  \end{eqnarray}  
The first term is the leading long time behaviour in (\ref{eq:h2-rho-zero-larget-leading}).
As before in the analogous short time expansion (\ref{eq:h2-rho-zero-smallt-gamma}), this long time gamma-resummed expansion (\ref{eq:h2-rho-zero-larget-gamma})  is convergent, in contrast to the long time asymptotic expansion (\ref{eq:h2-rho-zero-larget}). This leads to a dramatic improvement in the extrapolation from long time to shorter times, as illustrated in Figure \ref{fig:klarge-gamma}.

\section{Duality between short and long time asymptotics}
\label{sec:duality}

The asymptotic expansions (\ref{eq:h2-rho-zero-smallt}) and (\ref{eq:h2-rho-zero-larget}), and their gamma-resummed forms (\ref{eq:h2-rho-zero-smallt-gamma}) and (\ref{eq:h2-rho-zero-larget-gamma}), suggest the existence of a relation between the short time and long time asymptotics under the short time to long time duality transformation 
\begin{eqnarray}
t\to \frac{4\pi^2}{t}
\label{eq:tdual}
\end{eqnarray}
Note that the Dirichlet eta and Riemann zeta functions (which appear in these expansions) are related by a simple identity:
\begin{eqnarray}
 \eta(2n)=(1-2^{1-2n})\zeta(2n)
 \label{eq:eta-zeta}
 \end{eqnarray}
Writing this identity as $(1-2^{-2n}) \zeta(2n)=\frac{1}{2} \left(\eta(2n)+\zeta(2n)\right)$, we observe that the long time asymptotic expansion (\ref{eq:h2-rho-zero-larget}) can be written as 
\begin{eqnarray}
K(t, 0)&\sim & -\frac{e^{-t/4}}{4\pi^2 \sqrt{t}}\sum_{n=0}^\infty (-1)^n \eta(2n) \Gamma\left(n+\frac{1}{2}\right) \left(\frac{4}{t}\right)^n  
-\frac{e^{-t/4}}{4\pi^2 \sqrt{t}}\sum_{n=0}^\infty (-1)^n \zeta(2n) \Gamma\left(n+\frac{1}{2}\right) \left(\frac{4}{t}\right)^n
\quad, \quad t\to+ \infty
 \label{eq:split}
 \end{eqnarray}
Note that,  apart from the prefactor,  the first summation in (\ref{eq:split}) is the same as in the short time expansion (\ref{eq:h2-rho-zero-smallt}), with the replacement $t\to \frac{4\pi^2}{t}$.

 To formalize this duality transformation, it is convenient to define the normalized diagonal heat kernel $\tilde K(t, 0)$ by dividing out the leading short-time factor $e^{-t/4}/(4\pi t)$ (as has been done in the plots in Figures \ref{fig:kshort-expansions},  \ref{fig:kshort-gamma},
\ref{fig:klarge-expansions}, \ref{fig:klarge-gamma}):
 \begin{eqnarray}
 \tilde K(t, 0):=(4\pi t)e^{\frac{t}{4}} K(t, 0)
 \label{eq:ktilde}
 \end{eqnarray}
 Then the identity (\ref{eq:split}) can be written as a duality transformation
  \begin{eqnarray}
 \tilde K\left(\frac{4\pi^2}{t}, 0\right) &=&-\sqrt{\frac{4\pi}{t}}\tilde K(t, 0) + \sqrt{\pi t} \int_0^\infty dv\, v\, {\rm coth}\left(\pi v\right) e^{-\frac{v^2 t}{4}} 
 \label{eq:duality1}
 \end{eqnarray}
Replacing $t\to \frac{4\pi^2}{t}$, this can be equivalently expressed as 
 \begin{eqnarray}
 \tilde K\left(\frac{4\pi^2}{t}, 0\right)
= -\sqrt{\frac{\pi}{t}}\tilde K(t, 0) + \frac{2\pi^2}{t} \int_0^\infty dv\, v\, {\rm coth}\left(\pi v \right) e^{-\pi^2 v^2/t} 
 \end{eqnarray}
 From these integral representations we recognize that these duality transformations actually follow  from the simple trigonometric identity
 \begin{eqnarray}
\tanh\left(\frac{v}{4}\right)= -\frac{1}{\sinh\left(\frac{v}{2}\right)}+{\rm coth}\left(\frac{v}{2}\right)
\label{eq:trig}
 \end{eqnarray}
 
 Interestingly, the extra term in the duality transformation (\ref{eq:duality1}) is related to the diagonal propagator for a (Majorana) fermion on the hyperboloid  \cite{Camporesi:1992tm,Buchbinder:2014nia}:
 \begin{eqnarray}
K_{\rm spinor}\left(t, 0\right) &=&-\frac{1}{\pi} \int_0^\infty dv\, v\, {\rm coth}\left(\pi\, v\right) e^{-v^2 t} 
 \label{eq:spinor}
 \end{eqnarray}
 Thus, we can write the duality transformation  (\ref{eq:duality1}) as
  \begin{eqnarray}
 \tilde K_{\rm scalar}\left(\frac{4\pi^2}{t}, 0\right) &=&-\sqrt{\frac{4\pi}{t}}\tilde K_{\rm scalar}(t, 0)-\pi^{3/2}\sqrt{t}\, K_{\rm spinor}(t, 0)
 \label{eq:scalar-spinor}
   \end{eqnarray}

\section{Analytic continuation from heat kernel to Schr\"odinger kernel}
\label{sec:schr}

Since the short time and long time expansions of the heat kernel are asymptotic, 
the analytic continuation from the heat kernel to the Schr\"odinger kernel  (i.e., from $t\to i\, t$) requires some care. We illustrate this here for the diagonal kernels ($\rho=0$). The short time and long time Borel integral representations (\ref{eq:h2-rho-zero}) and (\ref{eq:h2-rho-large-d}) are exact and may be safely analytically continued with suitable contour deformation:
\begin{eqnarray}
K(i\, t, 0)
&=&\frac{i}{2}  \frac{e^{-i t/4}}{(4 \pi i t)^{3/2}}\int_{-\infty}^\infty
  ds\, \frac{s\, e^{- s^2/(4 t)}}{\sinh(e^{i\pi/4}\, s/2)}
  \label{eq:h2-short-it}\\
  &=& 
 -\frac{i}{2}  \frac{e^{-i t/4}}{2 \pi }\int_{-\infty}^\infty
  dv\,v\, \tanh(\pi \, e^{-i \pi/4}\,v)\,  e^{-v^2 t}
  \label{eq:h2-long-it}
  \end{eqnarray}
   \begin{figure}[htb]
 \centerline{\includegraphics[scale=.75]{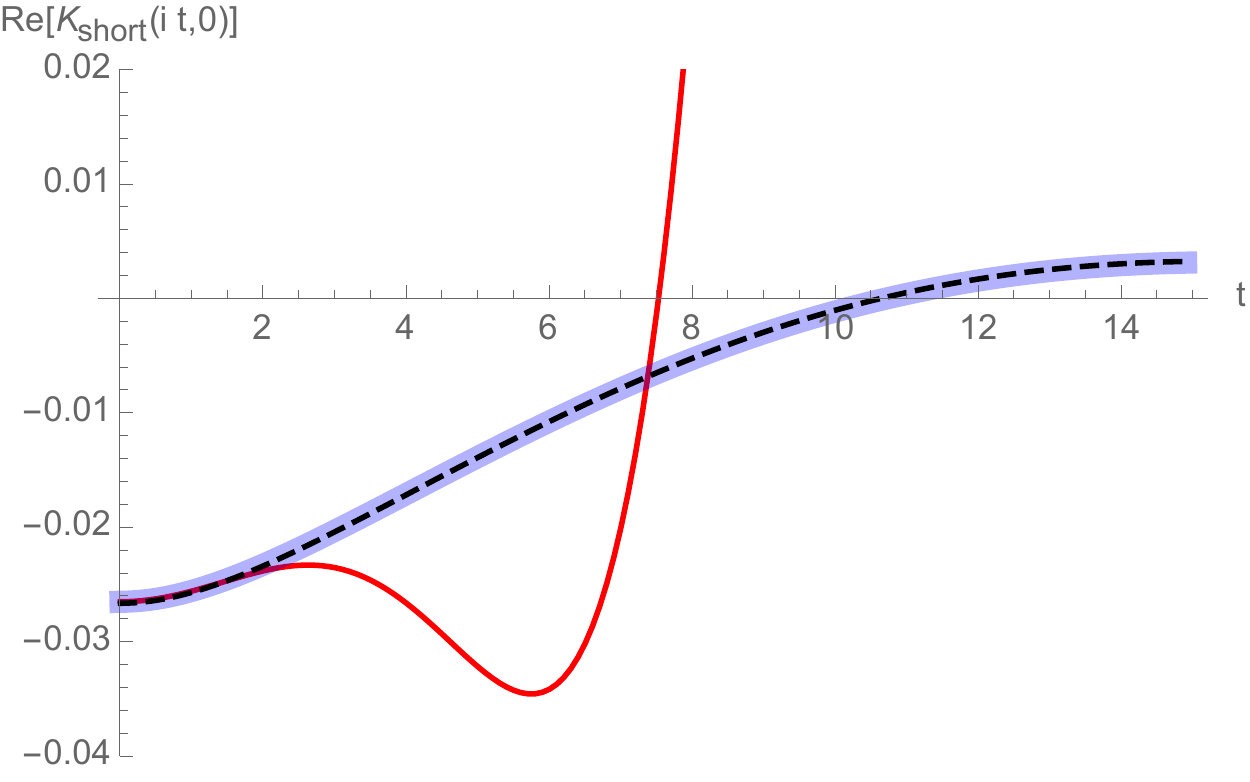}}
 \caption{The real part of $K_{\rm short}(i\, t, 0)$, the diagonal Schr\"odinger kernel at short time: the exact integral representation (\ref{eq:h2-short-it}) [thick blue curve], and the truncated short time asymptotic expansion (\ref{eq:short-truncated}), truncated at 5 terms [red curve]. The truncated short time gamma-resummed expansion (\ref{eq:short-truncated-gamma})  [black dashed curve], also truncated at 5 terms, is indistinguishable from the exact expression, out to large values of $t$.}
 \label{fig:kshort-imag}
 \end{figure}
   \begin{figure}[htb]
 \centerline{\includegraphics[scale=.75]{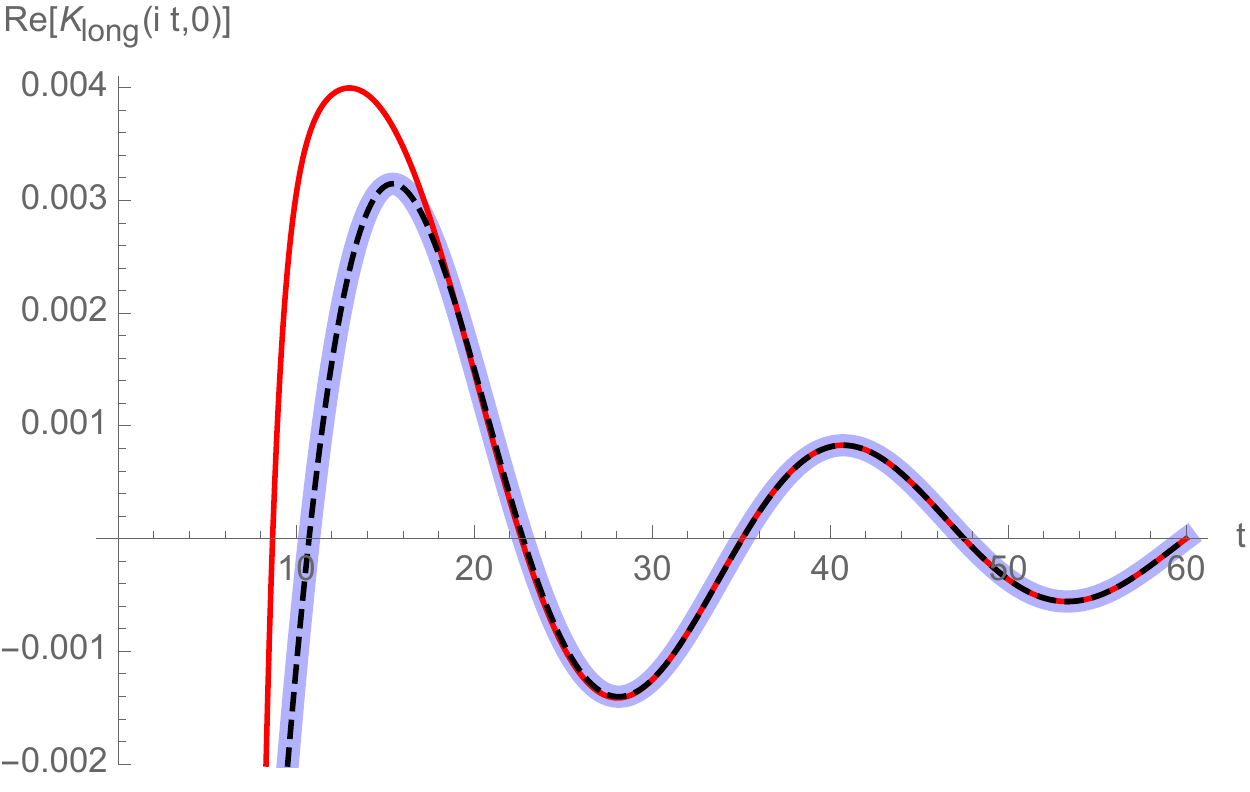}}
 \caption{The real part of $K_{\rm long}(i\, t, 0)$, the diagonal Schr\"odinger kernel at long time: the exact integral representation (\ref{eq:h2-long-it}) [thick blue curve], and the truncated long time asymptotic expansion (\ref{eq:long-truncated}), truncated at 5 terms [red curve]. The truncated long time gamma-resummed expansion (\ref{eq:long-truncated-gamma})  [black dashed curve], also truncated at 5 terms, is indistinguishable from the exact expression, down to much shorter times.}
 \label{fig:klong-imag}
 \end{figure}
 
On the other hand, {\it truncations} of the short time and long time asymptotic expansions (\ref{eq:h2-rho-zero-smallt}) and (\ref{eq:h2-rho-zero-larget}) lack global information,
 and therefore cannot be simply rotated from $t\to i t$. Such truncated expansions would be:
\begin{eqnarray}
K_{\rm short}^{\rm trunc} (i t, 0) :=  \frac{e^{-i t/4}}{2 \pi^{3/2} i t}  \sum _{n=0}^N  \eta (2 n)\, \Gamma \left(n+\frac{1}{2}\right)\, \left(\frac{-i t}{\pi^2}\right)^{n}
\label{eq:short-truncated}
\end{eqnarray}
\begin{eqnarray}
K_{\rm long}^{\rm trunc}(i t, 0) :=
\frac{16}{\sqrt{\pi }}  \frac{e^{-i t/4}}{(4 \pi i t)^{3/2}} \hskip -2pt \sum _{n=0}^N 
\hskip -3pt \left(1-2^{-(2n+2)} \right) \hskip -3pt \zeta (2 n+2) \Gamma \left(n+\frac{3}{2}\right)
\hskip -4pt \left(\frac{4 i}{t}\right)^n
\label{eq:long-truncated}
\end{eqnarray}

However, the gamma-resummed expansions (\ref{eq:h2-rho-zero-smallt-gamma}) and (\ref{eq:h2-rho-zero-larget-gamma}) provide an accurate analytic continuation when truncated.
Such truncated expansions would be:
\begin{eqnarray}
K_{\rm short, gamma}^{\rm trunc}(i t, 0) := \frac{e^{-i t/4}}{4 \pi i  t}  \left(1+\frac{\pi}{\sqrt{i t}}\sum_{k=1}^N  (-1)^k k\, e^{-i \pi^2k^2/t}\,\Gamma\left(-\frac{1}{2}, \frac{\pi^2 k^2}{i t} \right) \right)
\label{eq:short-truncated-gamma}
\end{eqnarray}
\begin{eqnarray}
K_{\rm long, gamma}^{\rm trunc}(i t, 0) &:= &
\frac{e^{-i t/4}}{(4 \pi i t)^{3/2} }  \left(\pi^2 
- 3(i t)^{3/2} \sum_{k=1}^N k\left[\frac{1}{2} e^{i k^2 t/4}\,\Gamma\left(-\frac{3}{2},\frac{i k^2 t}{4}\right)
- e^{i k^2 t}\,\Gamma\left(-\frac{3}{2}, i k^2 t \right)   \right]\right)
\label{eq:long-truncated-gamma}
\end{eqnarray}
These truncated gamma-resummed expansions (\ref{eq:short-truncated-gamma})--(\ref{eq:long-truncated-gamma}) are much more accurate than the corresponding truncated asymptotic expansions (\ref{eq:short-truncated})--(\ref{eq:long-truncated}) because the incomplete gamma functions encode the appropriate analytic properties.
This is illustrated in Figures \ref{fig:kshort-imag} and \ref{fig:klong-imag}.

\section{Analytic continuation from $\mathbb H^2$ to $\mathbb S^2$}
\label{sec:sphere}

The analytic continuation from heat propagation on the 2 dimensional hyperboloid $\mathbb H^2$ to the 2 dimensional sphere $\mathbb S^2$ is achieved by $t \to -t$ and $a^2\to -a^2$, where $1/a^2$ is the constant curvature of the space   \cite{Camporesi:1990wm}.  On the sphere the heat kernel is a function of the elapsed time $t$ and the angular parameter $\theta$ characterizing the geodesic distance.
On $\mathbb S^2$, the short time heat kernel is  \cite{Camporesi:1990wm}
\begin{eqnarray}
\mathcal K(t, \theta) =  \frac{\sqrt{2}\, e^{t/4}}{(4\pi t)^{3/2}} \sum_{n=-\infty}^\infty (-1)^n \int_\theta^\pi d\phi\,\frac{(\phi+2\pi n)}{\sqrt{\cos\theta-\cos \phi}} e^{-\frac{1}{4t}(\phi+2\pi n)^2}
\label{eq:sphere-short}
\end{eqnarray}
and the long  time heat kernel is 
\begin{eqnarray}
\mathcal K(t, \theta) = \frac{e^{t/4}}{4\pi} \sum_{n=0}^\infty  (2n+1) P_n(\cos\theta) e^{-\left(n+\frac{1}{2}\right)^2 t}
\label{eq:sphere-long}
\end{eqnarray}
The short time expression (\ref{eq:sphere-short}) is a geometric expansion in terms of geodesics on the sphere, while the long time expression (\ref{eq:sphere-long}) is a spectral expansion. The geodesic expansion on $\mathbb S^2$ can be traced to the {\it complex} geodesics on $\mathbb H^2$, associated with the complex Borel singularities in (\ref{eq:borel-small-poles}), depicted in Figures \ref{fig:small-t-borel} and \ref{fig:small-t-poles}.
  \begin{figure}[htb]
 \centerline{\includegraphics[scale=.75]{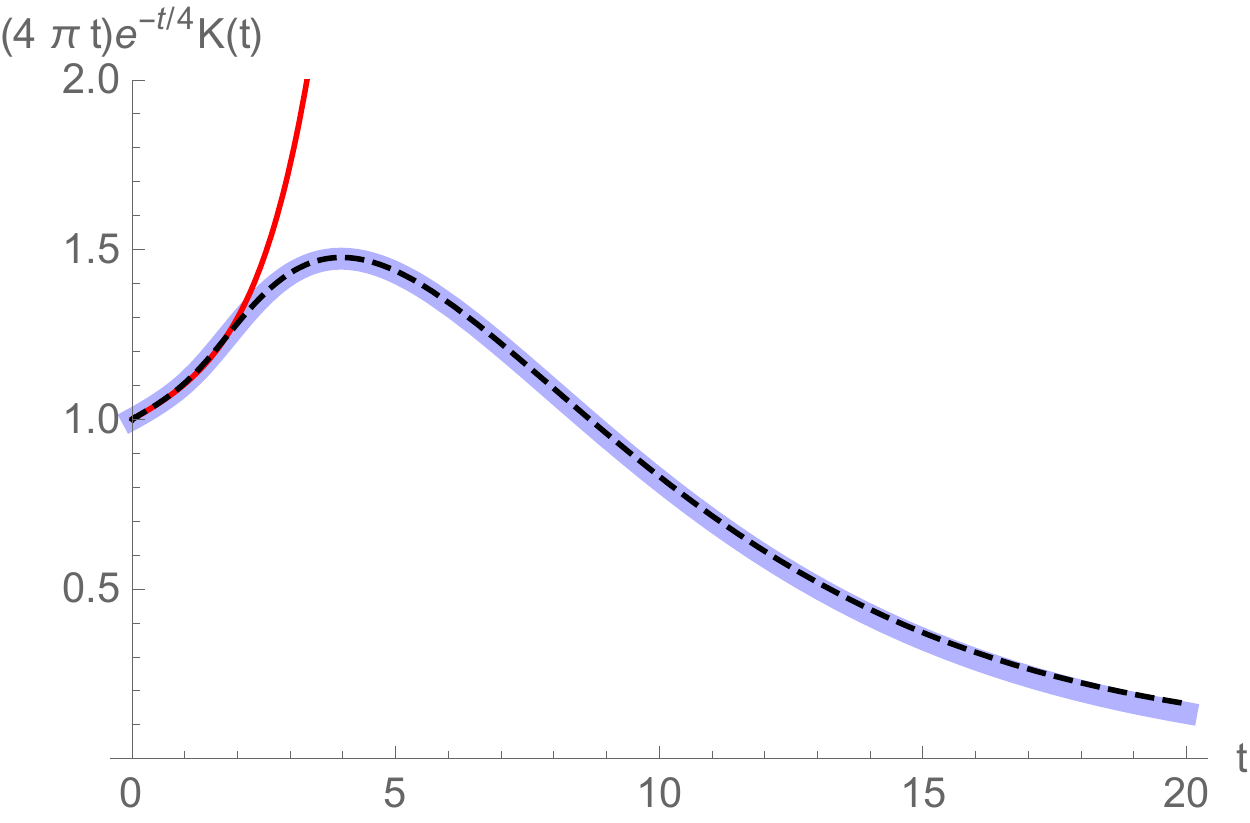}}
 \caption{Plot of $(4\pi t)\,e^{-t/4}\, {\mathcal K}_{\rm short}(t, 0)$, the normalized short time diagonal heat kernel (\ref{eq:sphere-short-diag}) on the two-sphere $\mathbb S^2$  (thick blue curve), compared with five terms (red curve) of the analytic continuation $t\to -t$ from  the short time asymptotic expansion (\ref{eq:h2-rho-zero-smallt}) of the diagonal heat kernel on the hyperboloid  $\mathbb H^2$. The black-dashed curve shows  five terms of the analytic continuation $t\to -t$ from the short time gamma-resummed expansion (\ref{eq:h2-rho-zero-smallt-gamma}) of the diagonal heat kernel on the hyperboloid  $\mathbb H^2$. This latter analytic continuation is much more accurate when we also extrapolate from short times to longer times.}
 \label{fig:k-sphere-short}
 \end{figure}
    \begin{figure}[htb]
 \centerline{\includegraphics[scale=.75]{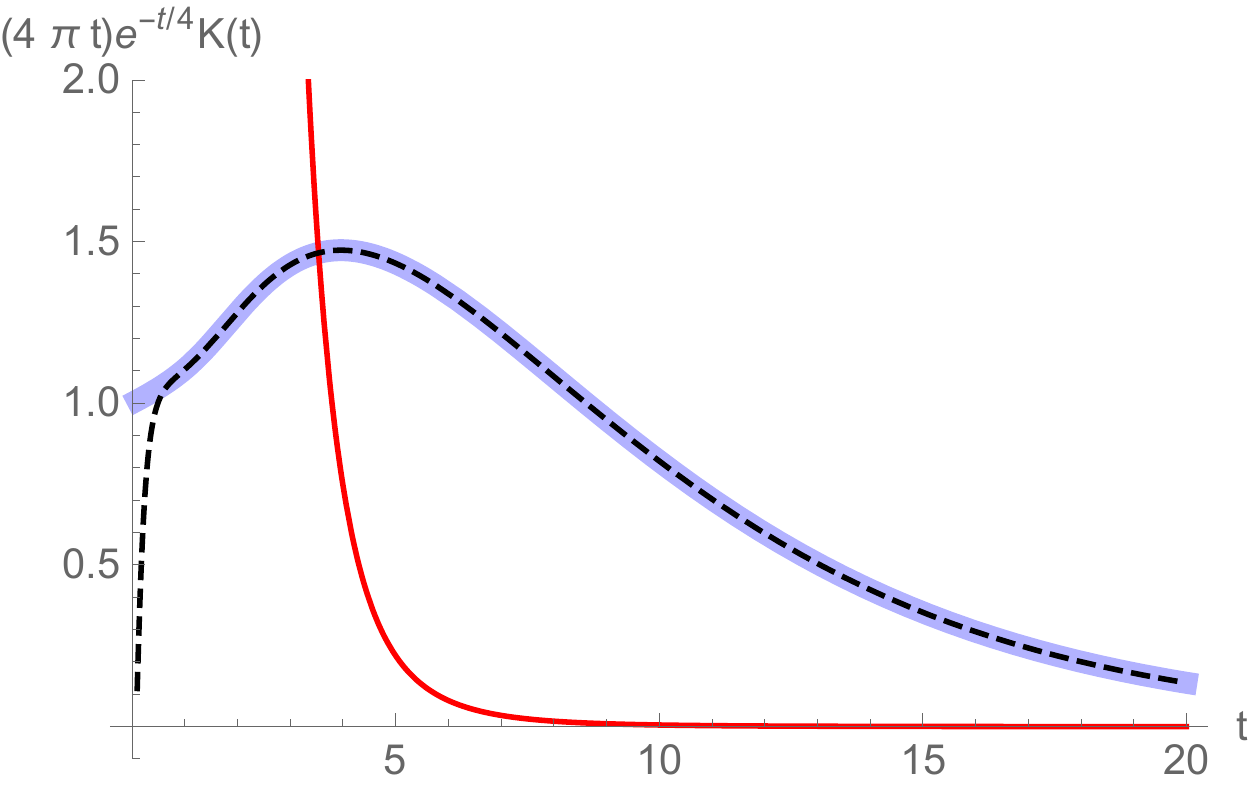}}
 \caption{Plot of $(4\pi t)\,e^{-t/4}\, {\mathcal K}_{\rm long}(t, 0)$, the normalized long time diagonal heat kernel (\ref{eq:sphere-long-diag}) on the two-sphere $\mathbb S^2$  (thick blue curve), compared with five terms (red curve) of the analytic continuation $t\to -t$ from  the long time asymptotic expansion  (\ref{eq:h2-rho-zero-larget})  of the diagonal heat kernel on the hyperboloid  $\mathbb H^2$. The black-dashed curve shows  five terms of the analytic continuation $t\to -t$ from the long time gamma-resummed expansion (\ref{eq:h2-rho-zero-larget-gamma})  of the diagonal heat kernel on the hyperboloid  $\mathbb H^2$. This latter analytic continuation is much more accurate when we  also extrapolate from long times to shorter times.}
 \label{fig:k-sphere-long}
 \end{figure}
     \begin{figure}[htb]
 \centerline{\includegraphics[scale=.75]{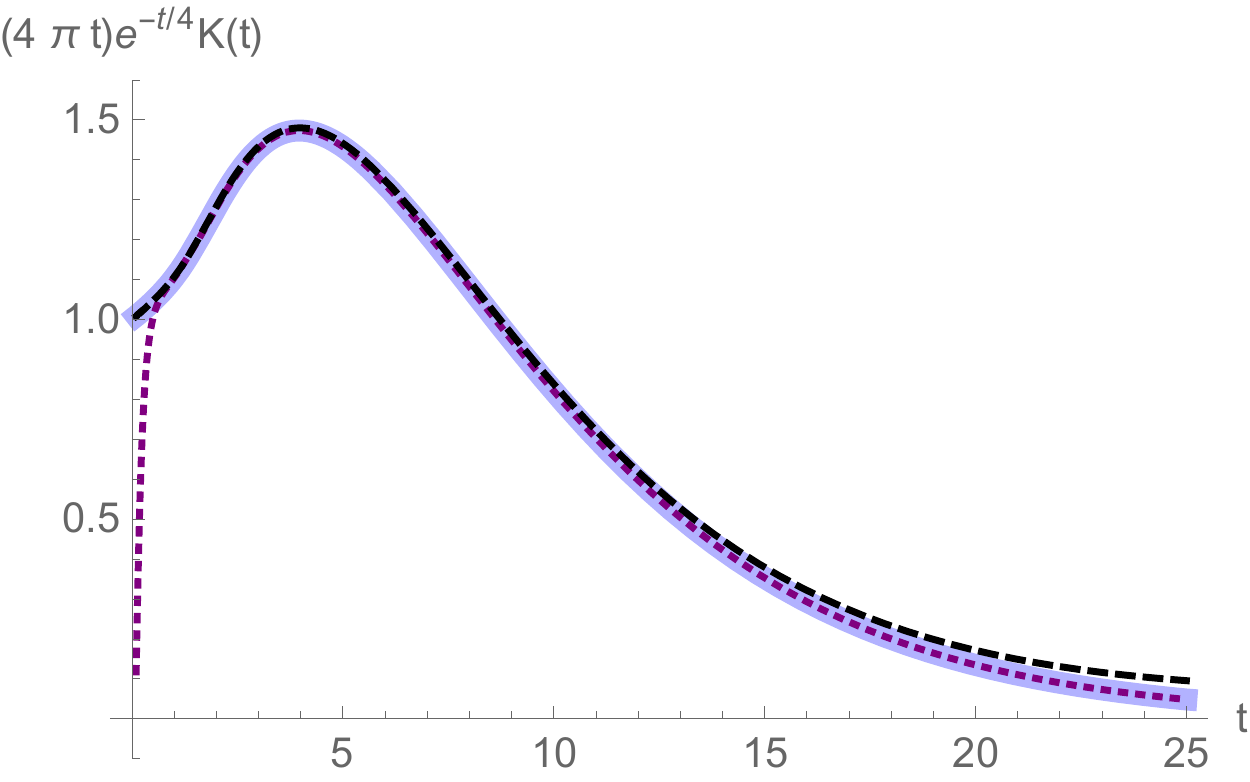}}
 \caption{Plot of $(4\pi t)\,e^{-t/4}\, {\mathcal K}(t, 0)$, the normalized diagonal heat kernel on the two-sphere $\mathbb S^2$  (thick blue curve), compared with five terms (purple dotted curve) of the analytic continuation $t\to -t$ of  the long time gamma-resummed expansion (\ref{eq:h2-rho-zero-larget-gamma})  of the diagonal heat kernel on the hyperboloid  $\mathbb H^2$. The black dashed curve shows  five terms of the analytic continuation $t\to -t$ of  the short time gamma-resummed expansion (\ref{eq:h2-rho-zero-smallt-gamma}) of the diagonal heat kernel on the hyperboloid  $\mathbb H^2$. The analytic continuations of these gamma-resummed expansions extrapolate very accurately between the extreme long and short times.}
 \label{fig:k-sphere-full}
 \end{figure}

The corresponding {\it diagonal} heat kernels arise when $\theta\to 0$.
\begin{eqnarray}
\mathcal K(t, 0) = \frac{e^{t/4}}{(4\pi t)^{3/2}} \sum_{n=-\infty}^\infty (-1)^n \int_0^\pi d\phi\,\frac{(\phi+2\pi n)}{\sin\left(\frac{\phi}{2}\right)} e^{-\frac{1}{4t}(\phi+2\pi n)^2}
\label{eq:sphere-short-diag}
\end{eqnarray}
\begin{eqnarray}
\mathcal K(t, 0) = \frac{e^{t/4}}{4\pi} \sum_{n=0}^\infty  (2n+1)  e^{-\left(n+\frac{1}{2}\right)^2 t}
\label{eq:sphere-long-diag}
\end{eqnarray}

Since the transition from $\mathbb H^2$ to $\mathbb S^2$ involves a rotation $t\to -t$,  this requires care because the short and long time expansions are asymptotic. Nevertheless, as with the continuation from heat kernel to Schr\"odinger kernel, we show here that the expansions (\ref{eq:h2-rho-zero-smallt-gamma}) and (\ref{eq:h2-rho-zero-larget-gamma}) in terms of incomplete gamma functions accurately encode the appropriate behavior.

The truncated short time and long time asymptotic expansions (\ref{eq:h2-rho-zero-smallt}) and (\ref{eq:h2-rho-zero-larget}) lack global information,
 and therefore cannot be simply rotated from $t\to - t$. Such truncated expansions would be:
\begin{eqnarray}
K_{\rm short}^{\rm trunc} (-t, 0) :=  -\frac{e^{t/4}}{2 \pi^{3/2} t}  \sum _{n=0}^N \eta (2 n)\, \Gamma \left(n+\frac{1}{2}\right) \left(\frac{t}{\pi^2}\right)^{n}
\label{eq:sphere-short-truncated}
\end{eqnarray}
\begin{eqnarray}
K_{\rm long}^{\rm trunc}(-t, 0):= \frac{16}{\sqrt{\pi }}  \, \frac{e^{t/4}}{(-4 \pi t)^{3/2}}  \sum _{n=0}^N
\left(1-2^{-(2n+2)} \right) \zeta (2 n+2) \Gamma \left(n+\frac{3}{2}\right) \left(\frac{4}{t}\right)^n
  \label{eq:sphere-long-truncated}
  \end{eqnarray}  
 These truncated asymptotic expansions are not very accurate, except in the extreme short or long time limit, respectively. See Figures \ref{fig:k-sphere-short} and \ref{fig:k-sphere-long}.

However, the gamma-resummed short time and long time expansions from (\ref{eq:h2-rho-zero-smallt-gamma}) and (\ref{eq:h2-rho-zero-larget-gamma}) provide an accurate analytic continuation when truncated.
Such truncated expansions would be:
\begin{eqnarray}
K_{\rm short, gamma}^{\rm trunc}(-t, 0) := -\frac{e^{ t/4}}{4 \pi  t}  \left(1+\frac{\pi}{\sqrt{-t}}\sum_{k=1}^N  (-1)^k k\, e^{-\pi^2k^2/t}\,\Gamma\left(-\frac{1}{2}, -\frac{\pi^2 k^2}{t} \right)\right)
\label{eq:sphere-short-truncated-gamma}
\end{eqnarray}
\begin{eqnarray}
K_{\rm long, gamma}^{\rm trunc}(-t, 0)&:=&
\frac{e^{t/4}}{(-4 \pi t)^{3/2} }  \left(\pi^2 
- 3(- t)^{3/2} \sum_{k=1}^N k\left[\frac{1}{2} e^{-k^2 t/4}\,\Gamma\left(-\frac{3}{2},-\frac{k^2 t}{4}\right)
- e^{-k^2 t}\,\Gamma\left(-\frac{3}{2}, -k^2 t \right)   \right]\right)
\label{eq:sphere-long-truncated-gamma}
\end{eqnarray}
These truncated gamma-resummed expansions are much more accurate because the incomplete gamma functions encode the appropriate analytic properties. See Figures \ref{fig:k-sphere-short} -\ref{fig:k-sphere-full}.

\section{Conclusions}
 
 In this paper we have studied the asymptotic nature of the heat kernel expansion on even-dimensional hyperbolic spaces at both short and long times. These properties can all be deduced from the heat kernel on two-dimensional hyperbolic space. At both short time and long time, the expansions are asymptotic and we have analyzed the singularity structure of the associated Borel transforms.  These Borel transforms are interestingly different. 
 We present resummations in terms of incomplete gamma functions which provide remarkably accurate interpolations and analytic continuations with respect to the time parameter $t$, and also continuations from  negative curvature  hyperbolic space to the corresponding   positive curvature spherical space. In addition, we have shown that the diagonal heat kernel has a duality between short and long times that mixes the scalar heat kernel and the spinor heat kernel.
These features are strongly suggestive of the resurgent \cite{ecalle,costin-book,maxim} nature of the heat kernel on such manifolds.

\section*{Acknowledgements}
 
This work is supported in part by the U.S. Department of Energy, Office of High Energy Physics, Award  DE-SC0010339.

\end{document}